\documentclass{article}

\usepackage{PRIMEarxiv}

\usepackage[utf8]{inputenc} 
\usepackage[T1]{fontenc}    
\usepackage{hyperref}       
\usepackage{url}            
\usepackage{booktabs}       
\usepackage{amsfonts}       
\usepackage{nicefrac}       
\usepackage{microtype}      
\usepackage{lipsum}
\usepackage{fancyhdr}       
\usepackage{graphicx}       
\graphicspath{{media/}}     

\title{An Online Ensemble Learning Model for Detecting Attacks in Wireless Sensor Networks
}

\author{
  Hiba Tabbaa\\
  LIPIM\\
 University Sultan Moulay Slimane  \\
  Ensa Khouribga \\
 \texttt{tabbaahiba90@gmail.com} \\ 
   \And
  Samir Ifzarne \\
   LIPIM\\
   University Sultan Moulay Slimane  \\
  Ensa Khouribga \\
  \texttt{sifzarne@gmail.com} \\
  \AND
  Imad Hafidi \\
  LIPIM\\
 University Sultan Moulay Slimane \\
 Ensa Khouribga \\
  \texttt{i.hafidi@usms.ma} \\
}
\date{}

\begin{document}
\maketitle

\begin{abstract}
	 In today's modern world, the usage of technology is unavoidable and the rapid advances in the Internet and communication fields have resulted to expand the Wireless Sensor Network (WSN) technology. A huge number of sensing devices collect and/or generate numerous sensory data throughout time for a wide range of fields and applications. However, WSN has been proven to be vulnerable to security breaches, the harsh and unattended deployment of these networks, combined with their constrained resources and the volume of data generated introduce a major security concern. WSN applications are extremely critical, it is essential to build reliable solutions that involve fast and continuous mechanisms for online data stream analysis enabling the detection of attacks and intrusions. In this context, our aim is to develop an intelligent, efficient, and updatable intrusion detection system by applying an important machine learning concept known as ensemble learning in order to improve detection performance. Although ensemble models have been proven to be useful in offline learning, they have received less attention in streaming applications. In this paper, we examine the application of different homogeneous and heterogeneous online ensembles in sensory data analysis, on a specialized wireless sensor network-detection system (WSN-DS) dataset in order to classify four types of   attacks: Blackhole attack, Grayhole, Flooding, and Scheduling among normal network traffic. Among the proposed novel online ensembles, both the heterogeneous ensemble consisting of an Adaptive Random Forest (ARF) combined with the Hoeffding Adaptive Tree (HAT) algorithm and the homogeneous ensemble HAT made up of 10 models achieved higher detection rates of 96.84\% and 97.2 \%, respectively. The above models are efficient and effective in dealing with concept drift, while taking into account the resource constraints of WSNs.
\end{abstract}


\section{Introduction}
Over the past decade, through the continued increasing growth of the Internet in the same vein, the amount of services that comes along with it has influenced almost every aspect of our human being's life. Rapid technological advances in microelectronics on one hand and wireless communication technologies on the other have resulted in the development of affordable, versatile and ubiquitous embedded sensor systems. 

 Internet of Things (IoT) is an innovation that allows communication between extensive variety of intelligent electronic devices and sensors. Wireless Sensor Networks (WSN) is another rapidly developing technology that is employed in IoT systems. WSNs have attracted the attention of researchers and research and development departments, and it will not be an overstatement to consider this technology as one of the most researched areas in the last decade due to their ease of deployment and their wide fields of real-time applications that differ based on their own objectives and specific constraints. The areas of WSNs applications are various including security and surveillance, home automations, health care services, flora and fauna, urban, critical military surveillance, environment monitoring, and so forth \cite{marriwala2012approach}. The WSN market was valued at \$46.76 billion in 2020 and is expected to reach \$123.93 billion by 2026, at a Compound annual growth rate (CAGR) of 17.64\% over the forecast period of 2021 – 2026. Therefore, the applications of WSN network are growing on a day-to-day basis in a considerable way. 

However, a WSN has several resource constraints that include a limited amount of energy, short communication range, low bandwidth, less processing capabilities and storage in each node. In many applications of WSNs, Sensor Nodes (SN) are deployed in remote, hostile and unattended locations, therefore, it is impractical to carry out maintenance on the nodes after installation. In fact, the energy consumption of the sensors plays an important role in the lifetime of the network and became the predominant performance criterion in this field. Additionally, in such an environment these SNs may be subject to disruptive and malicious actions that may outright damage the proper functioning of the network. Applications of WSNs require a high level of security to provide basic security requirements such as confidentiality, integrity, authentication, availability of the data traffic and battery life of the SNs \cite{osanaiye2018denial,salmon2013intrusion}. Making these applications invulnerable to different types of threats and attacks such as Blackholes, Sinkholes, Greyholes and so forth. These malicious attacks all cause the network traffic to deviate from the normal traffic for instance by interception of data sent / received by wireless medium and subsequently the ability to modify and replay the data. The intruder can also inject, saturate or damage network equipment. In critical applications, such attacks can be harmful and can cause major economic and security damage. 

There are different solutions that can be used to secure WSNs, such as key management, authentication or cryptography. Notwithstanding, these solutions do not guarantee complete prevention of all existing attacks. The most hardest challenge that the entire security sector faces is detecting and dealing with upcoming attacks. Howbeit, it is well known that intrusion detection systems (IDSs) are very effective security mechanisms to monitor the network from vicious attacks or unauthorized access as a second line of defense, and alert administrators on this subject \cite{abduvaliyev2013vital}. To summarize, IDS is needful to defend against WSN attacks.

The application of Machine Learning (ML) models in order to detect possible maliciousness in WSNs has largely increased in the last decade; however, the general approach in the literature still considers the analysis as an offline learning problem, where models are trained only one time on historical data. Because of the rising amount of data required to uncover increasingly sophisticated attacks, and given the large amount of data generated in real-time that gushes through these networks on a regular basis, traditional detection systems are inadequate for detecting malicious network intrusions. The  detection  of  attacks requires fast mechanisms for online  analysis  of thousands of events per second. This encourages the creation of a fast IDS for analyzing real-time network traffic determining instantaneously whether it is normal or exposed to any type of abnormal activities.

Stream machine learning consists of providing only a single sample (or a small batch of instances) to the learning algorithm at every time instant, with a very limited processing time, a finite amount of memory, and the necessity of having trained models at every scan of the streams. In addition, robust stream-based learning algorithms must be capable of detecting the drifts and updating their underlying models since a shift in data distribution (concept drift) can sometimes impact these streams of data, forcing the machine learning model to learn under non-stationary conditions. And yet, individual online learning methods are generally distinguished by a reduced detection rate. The second important requirement, aside from fast IDS, that should be considered when designing any IDS scheme for WSN:  IDS must have high accuracy and detection rate when detecting intruders \cite{mitrokotsa2008intrusion}.

Ensemble Learning (EL) approaches are based on the idea of gaining benefit from various classifiers by learning in an ensemble way. Since some classifiers may perform well for detecting a specific type of attacks and shows poor performance on other types. The EL works by building on strengths of various classifiers, through a combination of their results and then generating a majority vote out for classification. As a result, EL leads to maximizing accuracy through a reduction in variance and avoiding over-fitting \cite{opitz1999popular}.

 In this paper, we propose an ensemble stream-based machine learning approach for anomaly detection tailored to the WSNs characteristics.
  
\begin{enumerate}
\item\textit{ Contributions.} Our main contributions are summarized as follows:
\begin{itemize}
\item Evaluate the classification performance of multiple online individual algorithms such as k-nearest neighbor
(KNN), support vector machine (SVM), naive bayes (NB) and so on, under WSN for malicious intrusion detection.

\item Developing an investigation methodology to study the performance of different homogeneous ensemble approaches, such as hoeffding adaptive tree (HAT), and adaptive random forest (ARF),  along with heterogeneous ensembles based on two base-learners, suchlike ARF and NB. 

\item Examination of the ensemble performance with the existence of concept drift.
 
\end{itemize}

\item\textit{ Paper Organization.} The remainder of this paper is organized as follows: Section 2 presents related work in this field. Section 3 presents the proposed online intrusion detection scheme for WSN. Section 4 presents experimental environments of our study and section 5 analyzes the performance evaluation of the proposed approach. Finally, conclusions and future work are drawn in Section 6.
\end{enumerate}

\section{Related Work}

In defending against malicious attacks and misapprehensions in WSNs, various intrusion detection approaches have been proposed in the literature and they are mainly divided into anomaly detection, misuse detection, specification-based detection and hybrid system detection \cite{sen2020machine}. Recent researches are mostly concerned about anomaly-based IDS, thus our research focuses on this class. Anomaly-based IDS searches for both known and unknown patterns \cite{zamry2021lightweight}. Some credible anomaly detection approaches are currently provided based on the requirements of wireless sensor networks, notably including machine learning algorithms to create a classification model based on network traffic characteristics, artificial immune algorithms, clustering algorithms and statistical learning models.

Alqahtani \textit{et al.}\cite{alqahtani2019genetic} have proposed (GXGBoot) model to detect minority classes of attacks based on a genetic algorithm and an extreme gradient boosting (XGBoost) classifier, in highly imbalanced data traffics of wireless sensor networks. A set of experiments were conducted on (WSN-DS) dataset using held-out splitting and 10 fold cross-validation techniques. 10 fold cross-validation tests achieved a satisfactory results with high detection rates of 98.2\%, 92.9\%, 98.9\%, and 99.5\% for flooding, scheduling, grayhole, and blackhole attacks, respectively, in addition to 99.9\% for normal traffic.

Park \textit{et al.}\cite{park2018effective} compared random forest (RF) classifier with artificial neural network (ANN) algorithm for detecting the type of DoS attacks in WSNs, and it is found that the proposed RF classifier attains best F1-score results which are 96\%, 99\%, 98\%, 96\% and 100\% for flooding, blackhole, grayhole, scheduling (TDMA), and normal attacks, respectively. However, the outcome of this analysis was for a limited number of instances in the testing phase, which represents approximately 25\% (94,042 instances) of the results.

biswas \textit{et al.}\cite{biswas2021anomaly} introduced an anomaly detection strategy in WSNs utilizing ensemble random forest (ERF), with Decision Tree, Naive Bayes, and K-Nearest Neighbor as the ensemble's base learners. The random forest was also built using bootstrap sampling. The authors tested the ERF algorithm on a real-world sensor dataset, namely the activity identification based on multi-sensor data fusion (AReM) dataset.

Dong \textit{et al.}\cite{dong2020intrusion} proposed an intrusion detection model based on information gain ratio and Bagging algorithm for detecting DoS attacks in a cluster-based WSNs. To eliminate unnecessary features, the authors used the information gain ratio. The Bagging algorithm was used to build an ensemble algorithm which train a series of C4.5 decision trees in order to improve them. To test the model's accuracy, the proposed model was implemented using both the NSL-KDD and the WSN-DS datasets separately. 

Otoum \textit{et al.}\cite{otoum2020novel} proposed a novel methodology for detecting attacks in WSNs that use an ensemble classifier with Random Forest (RF), Density-Based Spatial Clustering of Applications with Noise (DBSCAN), and Restricted Boltzmann Machine (RBM) as basis classifiers. As a combination technique, Bayesian Combination Classification (BCC) has been used. For performance comparison, Independent BCC (IBCC) and Dependent BCC (DBCC) have been examined. The performance comparisons state that the ensemble technique DBCC-based IDS shows a promising result over the individual methods in attacks detection. 

Malmir \textit{et al.}\cite{malmir2019novel} proposed a novel ensemble approach for anomaly detection in WSNs using Time-overlapped Sliding Windows, evaluation results confirmed that the proposed method has a strong ability of attack classification and effectively improve the security system in terms of convenient metrics in the area of anomaly detection systems.

Kumari \textit{et al.}\cite{kumari2020hybrid}  developed an ensemble-based model for intrusion detection by combining these two machine learning techniques, J48 DT and SVM. The KDD99 intrusion detection dataset was optimized using particle swarm optimization to identify the nine most relevant and critical attributes, WEKA is utilized to implement classification. The suggested model yielded results with a higher accuracy of 99.1\% and a lower FAR of 0.9\%.

Fitni \textit{et al.}\cite{fitni2020implementation} proposed an ensemble-based AIDS model using DT, LR, and gradient boosting as inputs to an ensemble learning stacking classifier . The Chi-squared correlation approach was used to determine 23 relevant characteristics from the Communications Security Establishment and Canadian Institute for Cybersecurity 2018 (CSE-CIC-IDS2018) dataset. The proposed model surpasses seven individual classifiers with 98.8\% accuracy  and 97.1\% detection rate score.

However, none of the aforementioned works take into account continuous streaming data, and seldom work was done for anomaly detection for WSNs based on online ensemble learning in real-time. There are few works addressing anomaly detection in streaming data for embedded systems. 

The work in \cite{bosman2015ensembles} by authors bosman \textit{et al.} present a new lightweight architecture focused on ensembles of incremental learners for online anomaly detection in IoT applications, including WSNs. Also in environments with little a priori knowledge, their decentralized methodology outperformed each individual centralized offline learner alternatives in detecting anomalies determining that ensemble schemes are realistic to adopt. \\Ding \textit{et al.}\cite{ding2015novel} proposed a distributed online ensemble anomaly detector method in resource-constrained WSNs, ensemble pruning based on biogeographical based optimization (BBO) was employed to reduce the high resource demand and produce an optimized detector that performs at least as well as the original ones. The experiments operated on a real WSN dataset demonstrated the effectiveness of the proposed method.

Alrashdi \textit{et al.}\cite{alrashdi2019fbad} proposed a framework for identifying attacks in the fog node by employing the Online Sequential Extreme Learning Machine (OS-ELM) and majority voting to discover anomalies. The authors utilized the NSL-KDD to analyze and test their Framework. 

Ifzarne \textit{et al.}\cite{ifzarne2021anomaly} proposed an online learning classifier utilizing the information gain ratio to choose the relevant features of the sensor data with an online Passive aggressive algorithm in order to identify different types of DoS attacks. The (WSN-DS) dataset was utilized by the authors for the experiment.

Martindale \textit{et al.}\cite{martindale2020ensemble} proposed an approach for detecting intrusions in IoTs by exploring the performance and run-time trade-offs of a set of several online individual algorithms, as well as a few homogeneous and heterogeneous ensemble approaches. The massive online analysis (MOA) framework was used for implementing their approach. The 11 algorithms were run against three different KDDCup99 subsets. This study demonstrated that the ensembles outperformed the individual base learners, but at a higher cost in terms of run time, and the heterogeneous ensemble, which consisted of an ARF combined with HAT, outperformed the other online ensembles.

Although there are numerous studies exploring the use of online ensemble approaches and applying machine learning methods to streams of data, the majority of them ignore resource constraints and are targeted for Internet-of-Things (IoT) devices rather than WSNs.

\section{Research Methodology}

\subsection{WSN network topology based on LEACH routing protocol}

Many researchers used IDS to perform their work for WSN. Their work differs depending on the topology of the WSN and the protection method \cite{osanaiye2018denial}. WSN topologies are primarily divided into two types: flat based and cluster based. LEACH (Low Energy Adaptive Clustering Hierarchy) protocol is one of the main proactive sensor network protocols, and widely used clustering techniques in WSN. LEACH was proposed by Wendi B. Heinzelman of MIT \cite{xu2012improvement}. It is a self-organizing, hierarchical routing protocol based on adaptive clustering that is used in WSN to reduce the energy consumption of network elements in order to prolong their lifetime \cite{heinzelman2000energy} \cite{liu2006cluster}\cite{heinzelman2002application}. The LEACH consists of three parts: Sensor nodes (SN), cluster head (CH) nodes and base station (BS). The LEACH protocol's key concept is to group nodes into clusters in order to spread energy among all nodes in the network. The CH nodes gather and process the SN data in the cluster and transmit it to the base station. The cluster head nodes can monitor the behavior of the network traffic passing by in real time, and the intrusion detection model can be deployed as purely centralized where the IDS is installed in the base station. BS detects intrusions by analyzing the monitored network activity data.

  \begin{figure}[h]
  \begin{center}
    \includegraphics[height=2.5in,width=4in]{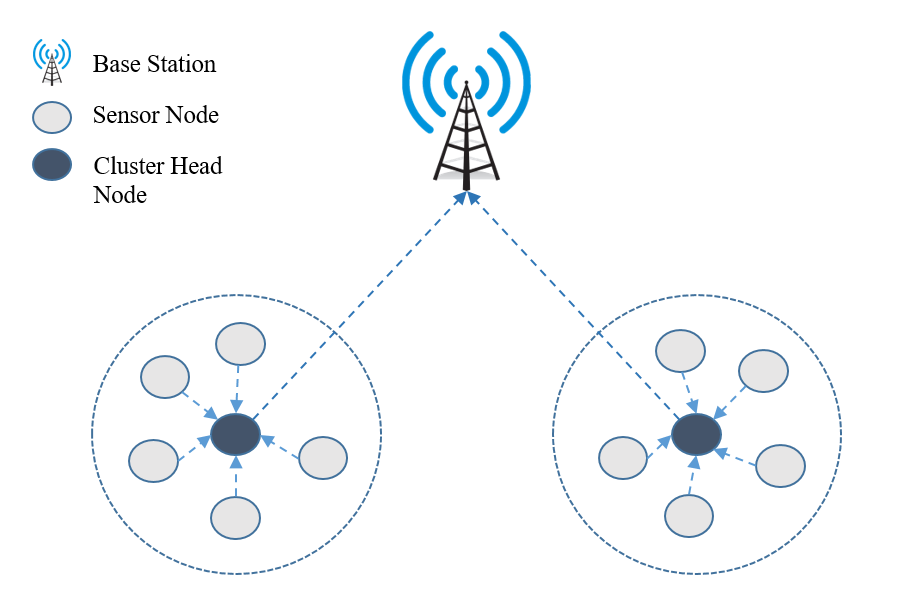}
  \end{center}
  \caption{WSN network topology.}
\end{figure}

\subsection{Classification Algorithms}

There are numerous models and algorithms for machine learning, the no free lunch (NFL) theorem stated by Wolpert et al. \cite{wolpert1997no} claims that in ML “there is no single learning algorithm that universally performs best across all domains” \cite{douglas2011performance}. The NFL theorem emphasizes the effectiveness of experimenting with various machine classifiers while tackling classification problems. Testing different classifiers is the most precise technique to solve domain specific problems, in our situation, the problem is intrusion detection. We will focus on two types of classification algorithms: single classifiers and ensembles.

\subsubsection{Single Classifier}
\begin{itemize}
\item \textbf{Support Vector Machine (SVM)}: Is a supervised learning algorithm that can be used for both linear and nonlinear problems. In SVMs, the idea is to find a max-margin separation hyperplane in the n-dimensional feature space.  Because the separation of hyperplanes is determined by a small number of support vectors, SVMs can yield satisfying results with small-scale training sets. On the other hand, SVMs are sensible to noise around the hyperplane. 
\item \textbf{k-Nearest Neighbors (KNN)}: Is a supervised learning algorithm that is very simplistic and can be used to fill in missing values and resample datasets. The core idea is to predict the class of a data sample using "feature similarity". In KNN algorithm the calculation of the distance from the neighbors to classify a sample based on its neighbors. The performance of KNN models is heavily influenced by the parameter k. As long as the value of K is very small, the more complex the model is and the higher the risk of overfitting. The larger k, on the other hand, the simpler the model and the lower the fitting ability \cite{ma2013k,cover1967nearest}. The value of K in our study is equal to 10.
\item \textbf{Naive Bayes (NB)}: Is a well-known classification technique, which is based on conditional probability and the concept of attribute independence that uses Bayes'\ theorem to forecast the likelihood or the  probability of an event occurring based on previous observations of related events \cite{kotsiantis2007supervised}. This algorithm is robust to noises and able to learn incrementally on the other hand the NB method performs poorly on attribute-related data.

\item \textbf{Passive Aggressive (PA)}: Is a family of online learning algorithms generally used for large-scale learning  that can be adopted for both classification and regression challenges. It is similar to the SVM classifier. The PA classifier attempts to find hyperplanes that can be used to split the instances into halves \cite{crammer2006online}.
\item \textbf{Perceptron (P)}: Is  a  supervised  learning  algorithm  for  classification and it is probably the most basic type of neural network model. The algorithm is made up of a single node or neuron that processes a row of data and predicts a class label.
\item \textbf{Hoeffding Tree (HT)}: An extremely fast decision tree technique for streaming data proposed by Domingos et al. \cite{domingos2000mining} in which we wait for new instances to arrive rather than reusing instances to compute the best splitting attributes. The HT's most intriguing characteristic is that it constructs a tree that provably converges to the tree constructed by a batch learner with suitably large data.

\item \textbf{Hoeffding Adaptive Tree (HAT)}: Adaptive tree monitors the performance of branches on the tree using adaptive windowing (ADWIN) algorithm \cite{bifet2007learning} to recent data and replaces them with new branches when their accuracy declines if the new branches are more accurate. 

\end{itemize}

\subsubsection{Ensemble Classifier}
When compared to a single model, ensemble learning enhances machine learning results by combining several models to improve predictive performance. There are two possibilities for an ensemble of classifiers:

\begin{itemize}
\item \textbf{Homogeneous Ensembles}: An ensemble of the same type of classifiers. A well-known example of this is random forests. A random forest is a homogeneous ensemble, which is a collection of many individual decision trees. 
\begin{enumerate}
\item Adaptive Random Forest (ARF):  A streaming classifier devoted for evolving data streams originally proposed in \cite{gomes2017adaptive}. ARF use a similar approach to the classical Random Forest algorithm \cite{breiman2001random}, in our approach the ARF was performed with the default settings which consisted of ten HTs algorithms as base learners and includes a drift detection operator. ARF made up of 20 HTs algorithms are tagged "ARF (20)".

\item Online Bagging: Authors in \cite{oza2001online,oza2001experimental} proposed the online bagging algorithm, which approximates the original random sampling with replacement by weighting instances giving each arriving example a weight according to Poisson(1).

\end{enumerate}

\item \textbf{Heterogeneous Ensembles:} An ensemble of different classifiers, which, for example, can contain SVM classifiers, neural network classifiers, and decision trees, all at the same time.
 
\begin{enumerate}
\item Weighted Majority (WM): A simple but widely studied algorithm proposed by \cite{littlestone1994weighted} that makes prediction based on a series of expert advices and learns to adjust their weights over time.
\end{enumerate}

\end{itemize}

\subsection{The proposed online ensemble intrusion detector model for WSNs}

The experiments in this work are designed to provide guidance on the appropriate ensemble technique for complying with the requirements of an ideal IDS for WSNs and this is accomplished by comparing the output of homogeneous ensembles composed of the same algorithm to build all the classifiers with those of heterogeneous ensembles composed of different algorithms. In our approach we propose online ensemble classification methods that attain higher detection rate in the aim of distinguishing malicious attacks while taking into account the resource constraints of WSNs.

 \begin{figure}[h]
 \centering
 \includegraphics[height=4in,width=6.5in]{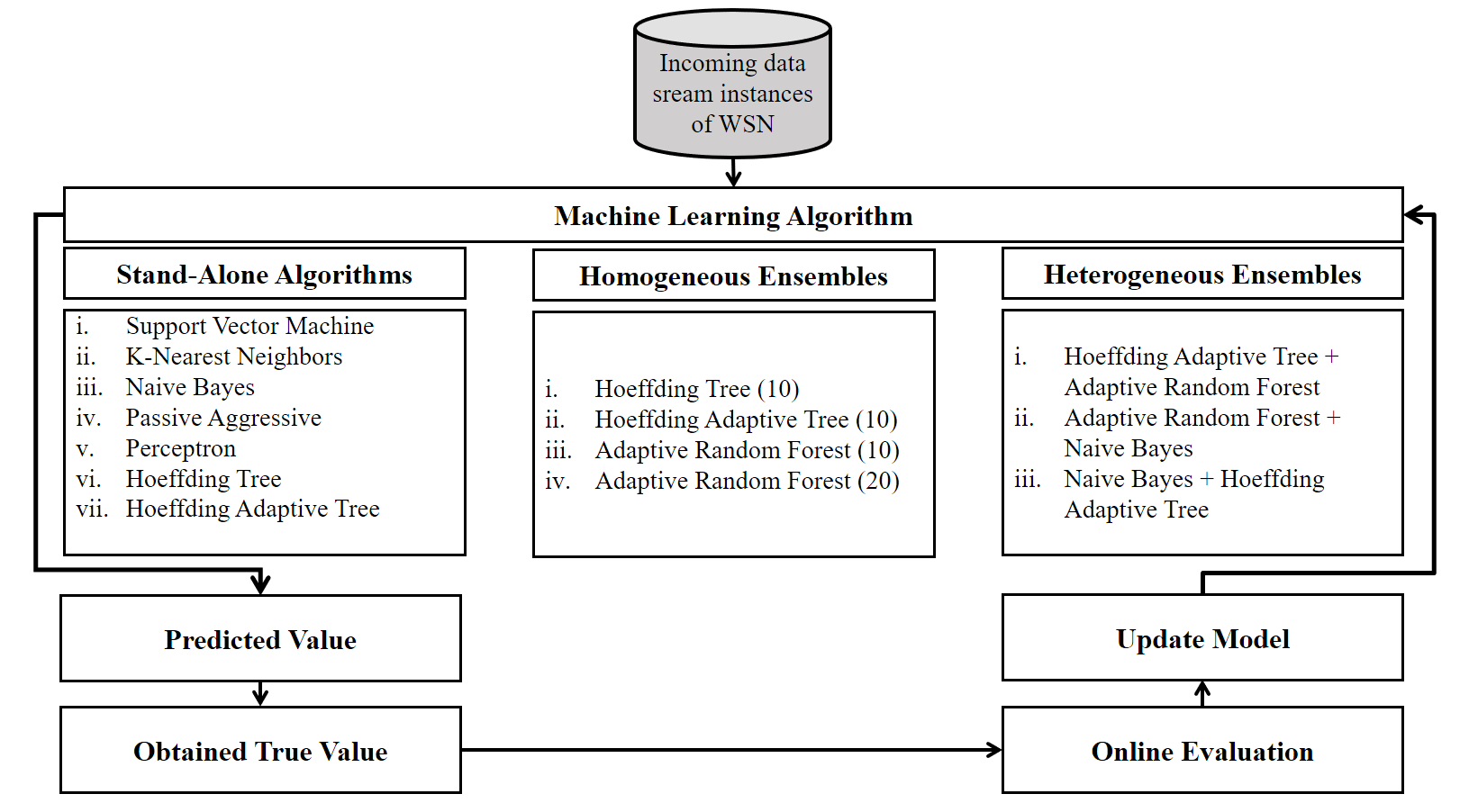}
 \caption{The structure of the proposed approach}
 \end{figure}
 
 Designing an ensemble often lies in two main challenges which are the choice of available base classifiers and the combiner methods.
\begin{itemize}
\item The detection of intrusions in our study consists of using online supervised learning and the performance is measured through using prequential (test-then-train approach) evaluation method, which was developed specifically for stream applications, where each sample is tested and then trained on in sequence by constructing a prediction with our current model for each incoming observation in the stream and score that prediction based on how well it matches the actual observation. The online model is then updated with the observation, and we proceed to the next instance. This evaluation technique is discussed in greater detail in the following section.

\item We have used (WSN-DS), a well-known dataset built exclusively for WSNs to improve DoS detection and classification which has been used in previous researches such as \cite{alqahtani2019genetic}, \cite{almomani2016wsn}, \cite{dong2020intrusion}, and we investigated the application of seven stand-alone algorithms, i.e. SVM, HAT, KNN, NB, PA, P, and HT to detect intrusions in WSNs.

\item Second, we inspected different homogeneous ensemble approaches, such as HAT and ARF using online bagging algorithm. The final online bagging model predicts by taking the simple majority vote of the predictions of the basis classifiers.

\item The performance of the ensemble is determined by two properties: The individual success of the ensemble's base classifiers and the independence of the base classifier's results from one another.
We compared numerous online classifiers to determine the best appropriate classifiers for online ensemble learning, and based on the results, the algorithms (HAT, ARF, NB) were chosen as base learners for the heterogeneous ensembles since they provided the highest predictive accuracy on streaming data. We employed a majority vote ensemble to combine the results of two different stable learners together in order to achieve both speed and precision at the same time. Three heterogeneous ensembles are proposed: HAT + ARF, ARF + NB, and NB + HAT.

\end{itemize}
\section{Experimental Environment}
\subsection{Experimental WSN-DS dataset }

In this work, we utilized the simulated wireless sensor network-detection system (WSN-DS) dataset developed by \cite{almomani2016wsn}, and the Network Simulator NS-2 was used to simulate the wireless sensor network environment \cite{issariyakul2009introduction}. Based on the LEACH routing protocol to gather the required data from the network with different attacking scenarios. The dataset consists of 23 features identifying the state of each sensor. LEACH protocol was selected because it is one of the most well-known energy efficient clustering algorithms for WSNs.The number of samples within the WSN-DS dataset is 374,661. Four types of Denial of Service (DoS) attacks are implemented in the constructed dataset: Blackhole, Grayhole, Flooding, and Scheduling attacks in addition to the normal behavior (no-attack) records. Using the label encoding technique, all of the categorical values of the label feature in the sample data are converted to numeric values to eliminate its impact on the algorithms. 

Simulation parameters are summarized in Table~\ref{table:WSNSimulation}. The distribution of the data is illustrated in Table~\ref{table:WSNDistribution}. Though, only 19 dimensional feature data involving the class label were in the dataset file as shown in Table~\ref{table:WSNFeatures}. The following are the technical characteristics of the computer used throughout the implementation phase:

\begin{itemize}
\item Central Processing Unit: Intel(R) Core(TM) i7-4610M CPU @ 3.00GHz 3.00GHz
\item Random Access Memory: 8 GB.
\item Operating System: Windows 7 Pro 64-bit.
\end{itemize}

\begin{table}[!htp]
\centering
\caption{WSN Simulation parameters.}
\label{table:WSNSimulation}
\begin{tabular}[t]{lc}
 \hline
Parameter & Value  \\
 \hline \hline
Number of nodes &100 nodes\\
Number of clusters &5\\
Network area &100m$\times$ 100m\\
Base station location  &(50,175)\\
Size of packet header  &25 bytes\\
Size of data packet  &500 bytes\\
Maximum transmission range  & 200 m\\
MAC protocol&CSMA/TDMA\\
Routing protocol  &LEACH\\
Simulation time  &3600 s\\
 \hline
\end{tabular}
\end{table}

\begin{table}[!htp]
\centering
\caption{Distribution of WSN-DS dataset.}
\label{table:WSNDistribution}
\begin{tabular}[t]{lcc}
 \hline
Types of attack & Quantity & Proportion(\%)  \\
 \hline \hline
Normal & 340066 & 90.77\\
Grayhole & 14596 & 3.90 \\
Blackhole  & 10049 & 2.68\\
Scheduling   &6638 & 1.77\\
Flooding  &3312 & 0.88\\
 \hline
\end{tabular}
\end{table}

\begin{table}[ht]
\centering
\caption{Features of the WSN-DS Dataset.}
\label{table:WSNFeatures}
\scalebox{0.80}{
\begin{tabular}[t]{cccc}
 \hline
Feature number & Symbol & Feature name  & Description \\
 \hline \hline
1 & id & Node Id& A unique ID number of the sensor node \\
2 &  Time & Time & The run-time of the node in the simulation\\
3 & Is\_CH & Is CH &  Describes if the node is a CH or not\\ 
4  & Who\_CH & Who CH & Cluster head ID\\
5  & Dist\_To\_CH & Distance to CH  & Distance between node and CH \\
6 & ADV\_S & ADV CH sends & Number of the advertise CH's broadcast messages sent to nodes\\
7  & ADV\_R & ADV CH receives & Number of advertise messages received by the nodes from CH\\
8  & JOIN\_S & Join request send & Number of join request messages sent by the nodes to the CH \\
9  & JOIN\_R & Join request receive & Number of join request messages received by CH from nodes\\
10  & SCH\_S & ADV SCH sends & Messages of TDMA  schedule broadcast sent to the nodes\\
11  & SCH\_R &  ADV SCH receives & Number of scheduled messages received by the CH\\
12  & Rank & Rank & Node order in TDMA scheduling \\
13  & DATA\_S & Data sent & Number of data packets sent from the node to its CH\\
14  & DATA\_R & Data received & Number of data packets received by the node from the CH\\
15  & Data\_Sent\_To\_BS & Data sent to BS & Number of data packets that are sent
from node to the BS\\
16  & dist\_CH\_To\_BS & Distance CH to BS & Distance between CH and BS \\
17  & send\_code & Send code &  The sending code of the cluster\\
18  & Consumed\_Energy & Energy consumption & Energy consumed\\
19  & Attack type & Attack type & Type of attacks or normal traffic\\
 \hline
\end{tabular}}
\end{table}

\subsection{The scikit-multiflow framework}

The experimental setup of this research aims to provide reproducibility, allowing different researchers to achieve with a high degree of agreement the same results obtained in this experiment by using the same experimental framework. However, the current experimental machine learning tooling is mainly divided between Java-based and Python-based implementations. Some researchers implement their experiments using tools built around Weka \cite{holmes1994weka} or the data stream mining framework MOA \cite{bifet2010moa}. Others prefer using solutions from the scikit-learn environment or the multioutput streaming platform scikit-multiow \cite{montiel2018scikit}.

 Scikit-mutliflow is a Python framework for learning from data streams and multi-output/multi-label learning. Scikit-multiflow is based on well-known open source frameworks like such scikit-learn, MOA, and MEKA. It includes techniques for classification, regression, concept drift detection, and anomaly detection. It also has a collection of data stream generators and evaluators. scikit-multiflow is intended to work with Python's numerical and scientific libraries NumPy and SciPy, as well as Jupyter Notebooks.
 
 \subsection{Evaluation technique}
\qquad
Considering the unbounded real-world streams of non-stationary data, classic techniques for evaluating the model on batch data such as train-test split and cross-validation
do not apply to models trained on streamed data \cite{gama2009issues}. To
succeed in dealing with this problem and obtain accurate measurements
over time we used the widely known prequential evaluation method in our experiment.

\qquad \textbf{The Prequential} evaluation method or interleaved test-then-train method. Each individual sample is used to test the model, which means to make a predictions, and then the same sample is used to train the model, and from this the accuracy can be incrementally updated. When the evaluation is intentionally performed in this order, the model is always being tested on instances that it has not seen yet. This approach has a favorable condition that no holdout set is needed for testing, making maximum use of the available data.

 \subsection{Evaluation metrics}
\qquad
Adequacy of the wireless sensor network intrusion detection algorithms were measured using the following measures: Accuracy (Acc), precision (P), recall (R), F1-Score (F), as well as the total running time (Training Time + Testing Time) in seconds (s) of the classification algorithm are collected and compared. Table~\ref{table:Metrics} shows the definitions of TP, FP, TN and FN.

An abnormal flow is treated as Positive (P) and normal flow is treated as Negative (N).
\begin{enumerate}
\item   \textbf{True Positive}: The model correctly predicts the positive class. The model correctly predicts that an instance is an attack by the classifier.
\item     \textbf{True Negative}: The model correctly predicts the negative class. The model correctly predicts that an instance is normal.
\item     \textbf{False Positive}: The model incorrectly predicts the positive class. The data samples (attack) predicted incorrectly as normal.
\item     \textbf{False Negative}: The model incorrectly predicts the negative class. The data samples (normal) predicted incorrectly as attack.
\end{enumerate}

  \begin{table}[ht!]
\centering
\caption{Definition of TP, FP, TN and FN.}
\label{table:Metrics}
\resizebox{\columnwidth}{!}{\begin{tabular}[t]{ccc}
\hline
 & Positive real class (Abnormal) &  Negative real class (Normal)\\
\hline\hline
Positive predicted class (Abnormal) & True Positive (TP) & False Negative (FN)\\
\hline
Negative predicted class (Normal) & False Positive (FP) & True Negative (TN)\\
\hline
\end{tabular}}
\end{table}%

These performance evaluation metrics are calculated as follows:

\begin{itemize}
\item \textbf{\textit{Accuracy (Acc):}} Represents the percentage of instances correctly classified. This is the number of correct predictions divided by the total number of predictions. Accuracy alone is not sufficient as a measure of performance, especially for datasets with unbalanced classes, it can be written as :
\begin{equation}
Accuracy = \frac{TP + TN}{TP + FP + FN + TN }
\end{equation}

\item \textbf{\textit{Precision (P) :}} Or Positive Predictive value (PPV) represents how good the model is at assigning positive events to the positive class. That is, how accurate the attack prediction is, and it is defined as follows:
\begin{equation}
Precision = \frac{TP }{TP + FP }
\end{equation}

\item \textbf{\textit{Recall (R) :}} It is also called as True Positive Rate (TPR), sensitivity and detection rate, measures how good the model is in detecting the positive class. So, given that attacks are the positive class, it represents the percentage of actually attacks identified correctly. Equation (3) represents the formula for calculating recall:
 
\begin{equation}
Recall = DetectionRate = \frac{TP }{TP + FN }
\end{equation}

\item \textbf{\textit{F1-Score (F) :}} Or F1-Measure represents is the harmonic mean of Precision and Recall i.e. Compare to the accuracy, f1-score is the best metric to check effectiveness of intrusion detection algorithm when IDS model uses unbalanced dataset and we search for a balance between precision and recall.
\begin{equation}
F1-Score = \frac{2 \times P \times R }{P + R}
\end{equation}
\end{itemize}

\section{EXPERIMENTAL RESULTS}
\subsection{Individuals Algorithms}
\begin{figure}[hbt!]
 \centering
 \includegraphics[width=6.5in,height=4in]{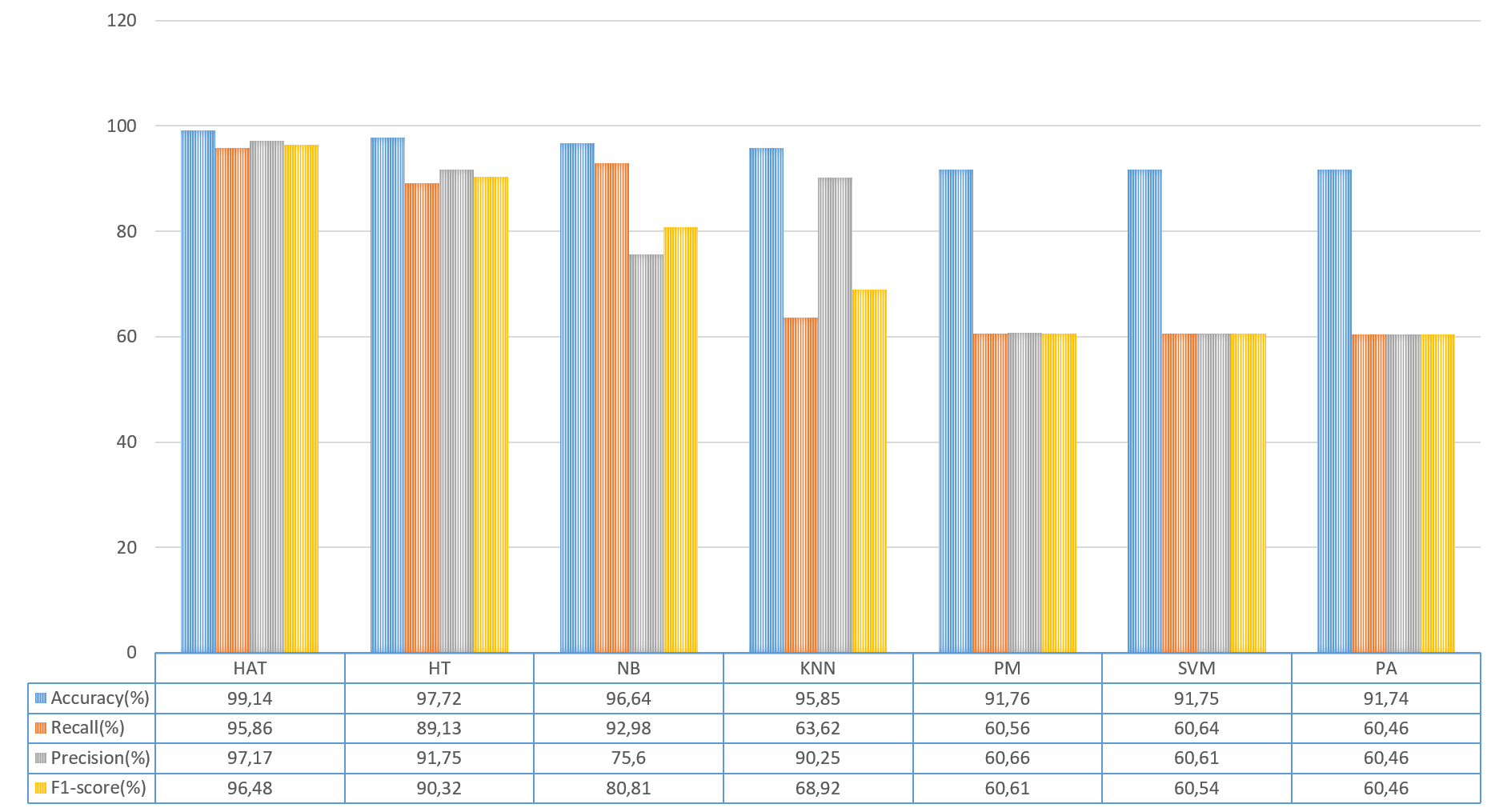}
 \caption{Comparison of performance of different individual models results.}
 \label{fig:singlealgo}
 \end{figure}
 This subsection presents and discusses the performance evaluation results of each algorithm when running individually. As can be seen from Figure~\ref{fig:singlealgo}, the R of HAT method reaches 95.86\% which is the highest detection rate compared with other online methods. The Acc of this method is 99.14\%, and for the P and F are respectively 97.17\% and 96.48\%. In second place, Naive Bayes with a detection rate of 92.98\% following by the HT method with a R of 89.13\%.
In summary, the HAT method performs better than the other online intrusion detection methods.

\begin{table}[hbt!]
\centering
\caption{Running time (Training Time + Testing Time) in seconds (s).}
\label{table:RunningtimeSing}
\begin{tabular}[t]{cc}
\hline
Models & Run-time (s)\\
\hline\hline
KNN & 1089.67\\
PA & 881.58 \\
PM & 785.87\\
SVM & 781.18\\
HAT & 218.08\\
NB & 141.95\\
HT & 100.28\\
\hline
\end{tabular}
\end{table}%

Comparing the model total running time, we can see that the HT had the fastest run-times followed by the model NB and in the third place the HAT method has a lower time than that of KNN, PM, SVM, and PA. In the other hand, the KNN model exhibits considerably a longer running time with approximately 18 minutes compared with other tested methods. The necessity to continuously calculate the distance between the predict target and every other sample still in memory by the KNN algorithm might give an explanation for the long running-time. When it comes to some critical streaming applications, such as a predictive model for network security intrusion, a fast model but less accurate is often preferred over a slow and more accurate model.

\subsection{Ensemble Algorithms }
\subsubsection{Homogeneous ensembles}
\begin{figure}[hbt!]
 \centering
 \includegraphics[width=5.5in,height=3.5in]{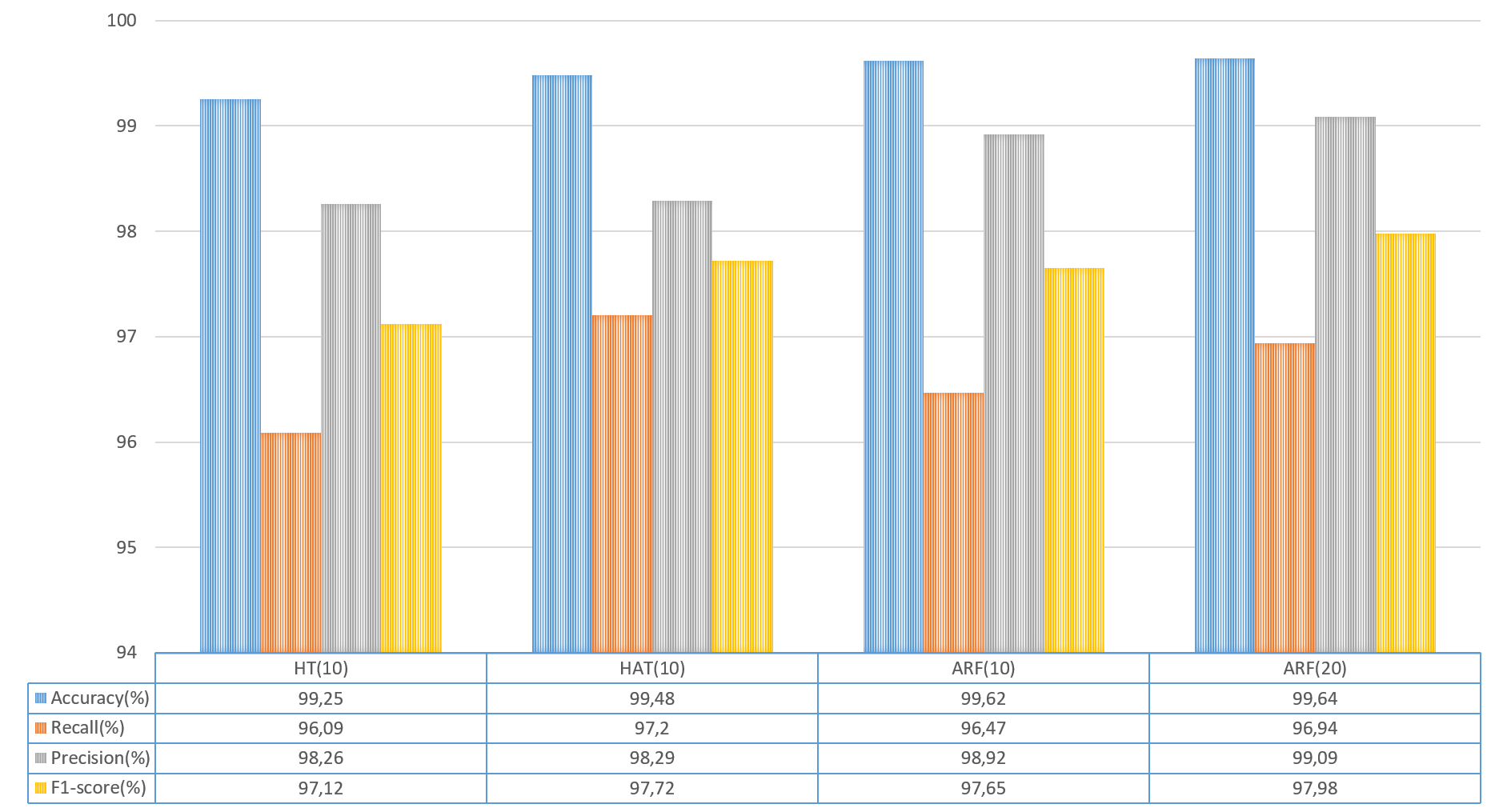}
 \caption{Homogeneous ensembles results. }
  \label{fig:homoalgo}
 \end{figure}

  Figure~\ref{fig:homoalgo} displays diverse homogeneous ensemble approaches, where each ensemble is made up of multiple instances of the same base learner. All of the ensemble approaches that have been tested have offered outstanding results, and especially HAT(10) as it is apparent, with a detection rate (R) that reaches 97.2\%. Immediately afterwards, ARF(20) with a R that attains 96.94\%. When comparing Figure~\ref{fig:singlealgo} with Figure~\ref{fig:homoalgo} in terms of results concerning HT, HT(10), HAT and HAT(10), as indicated by the homogeneous ensembles results, combining numerous online learnables prediction enhances the detection of attacks greatly when compared to single predictors. Increasing the amount of trees substantially enhance performance as it is the case when comparing the results of AFR(10) and AFR (20). 
 
\begin{table}[hbt!]
\centering
\caption{Running Time metric in seconds (s).}
\label{table:Runningtimehomo}
\begin{tabular}[t]{cc}
\hline
Models & Run-time (s)\\
\hline\hline
HT(10) & 971.95\\
HAT(10) & 2400.01 \\
ARF(10) & 2049.89\\
ARF(20) & 4214.41\\
\hline
\end{tabular}
\end{table}%

 We can observe from Table~\ref{table:Runningtimehomo} that HT(10) is faster than the rest of online homogeneous ensembles and ARF(20) has a higher classification time. Our results highlight that the running time of the ensemble algorithms are significantly longer than the individual methods when comparing Table~\ref{table:RunningtimeSing} with Table~\ref{table:Runningtimehomo}. The number of estimators does have an impact on the model's running time.

\subsubsection{Heterogeneous ensembles}
\begin{figure}[h]
 \centering
 \includegraphics[width=6in,height=4in]{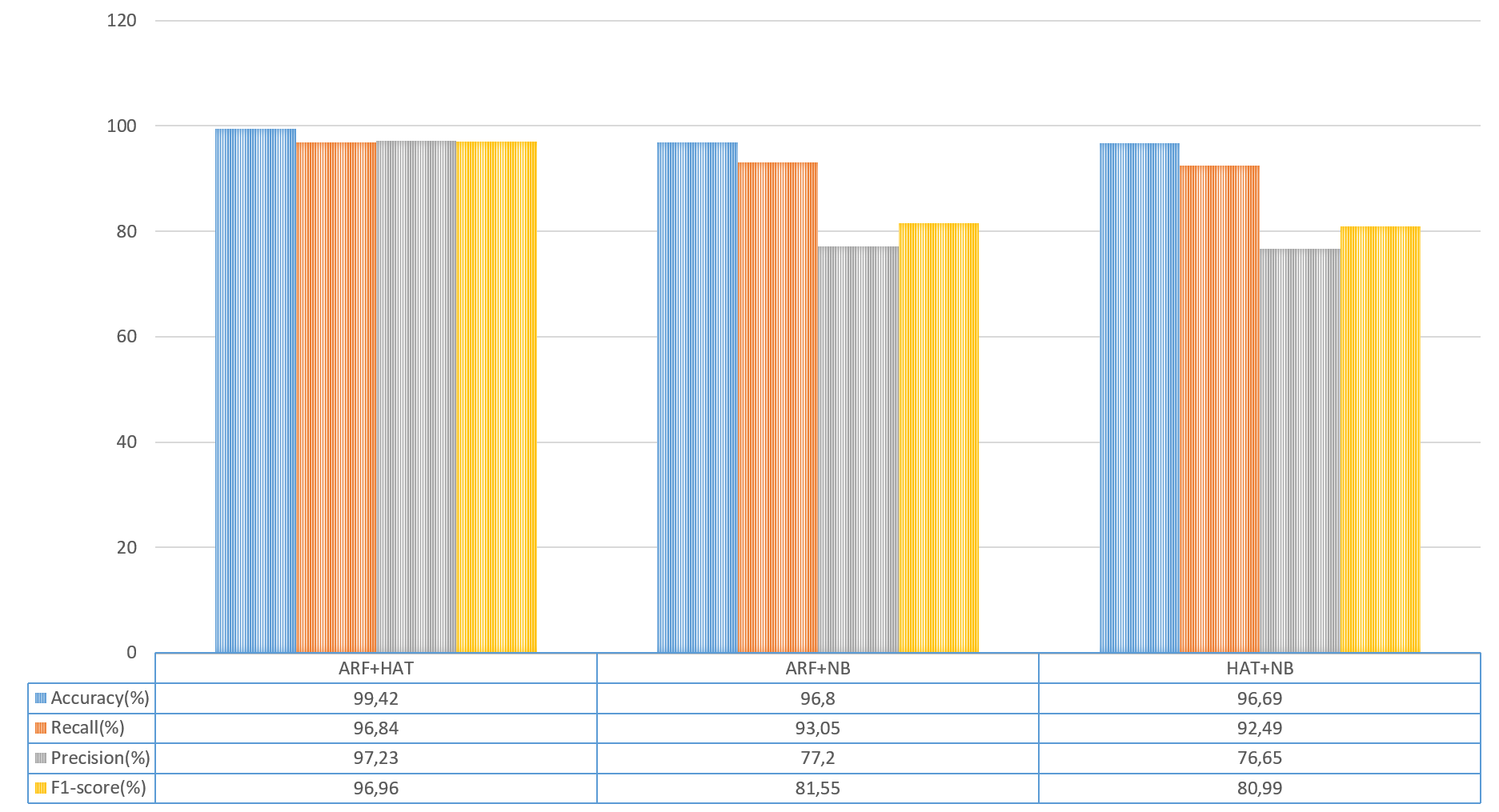}
 \caption{Heterogeneous ensembles results. }
 \label{fig:heteroalgo}
 \end{figure}

  Restating that the ARF was run with the default settings, which included 10 HTs algorithms. As can be seen from Figure~\ref{fig:heteroalgo} which presents the predictive performance of the heterogeneous ensembles, the online evaluation results show that the detection rate of the method (ARF + HAT) in this paper reaches 96.84\%, which is higher than that of (ARF + NB) and (HAT + NB). The Acc, P and F are respectively 99.42\%, 97.23\% and 96.96\%. According to the performance comparisons of ARF with R of 96.47\% and HAT with R of 95.86\%, the heterogeneous ensemble of (ARF+HAT) methodology outperforms the seven stand-alone algorithms and the homogeneous ensemble of ARF(10) in overcoming the misclassification of attacks, the reason is that the accuracy of individual models and diversity among individual models are all aspects that contribute to the heterogeneous ensemble's success.
 
  \begin{table}[ht]
\centering
\caption{Running Time metric in seconds (s):}
\label{table:Runningtimeheteroge}
\begin{tabular}[t]{cc}
\hline
Models & Run-time (s)\\
\hline\hline
ARF+HAT & 2345 \\
ARF+NB & 2192 \\
HAT+NB & 403 \\
\hline
\end{tabular}
\end{table}%
 
  We can see from Table~\ref{table:Runningtimeheteroge} that the heterogeneous ensemble of (HAT + NB) technique has a relatively short running time as it contains two base learners. On the other hand, the (ARF + HAT) and (ARF + NB) classification times are close to each other, with the (ARF + HAT) method being more time-consuming with correspondingly highest detection rate. 
  
\subsubsection{Concept drift results }

\begin{figure}[h]
 \centering
 \includegraphics[width=6in,height=3.5in]{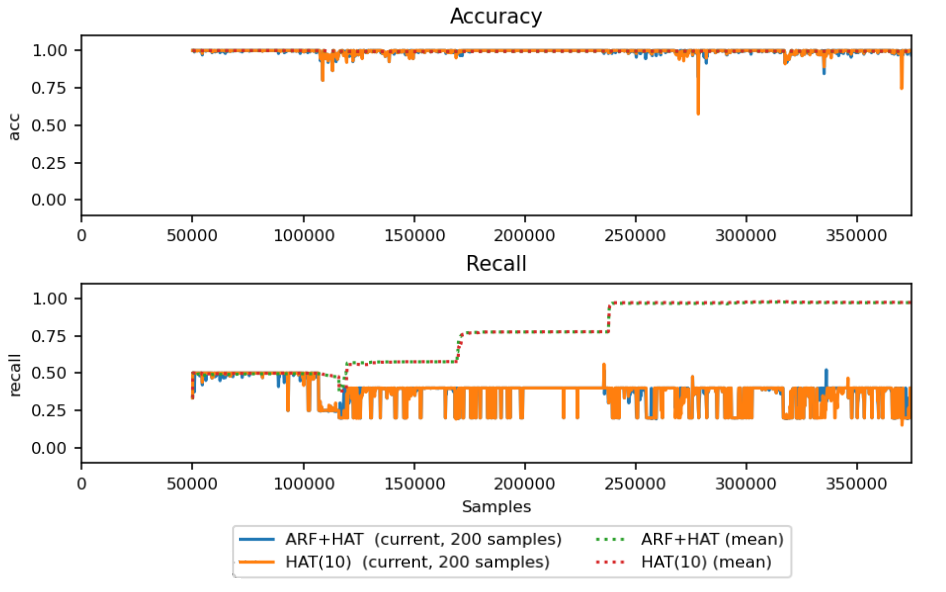}
 \caption{Concept drift results. }
 \label{fig:conceptdrift}
 \end{figure}

The existence of concept drift in streaming data is a significant element that contributes to a decrease in predictive accuracy over time. Figure~\ref{fig:conceptdrift} depicts a plot of the predictive accuracy offered over time for the homogeneous ensemble HAT(10) as well as the heterogeneous ensemble ARF + HAT presented by (orange) and (blue) respectively both of which performed favorably in terms of accuracy and classification of attacks. In terms of concept drift, however, there is no evident performance advantage between these two. While the HAT(10) ensemble's classification performance suffers from significant dips in our results, there are several small areas where the HAT (10) ensemble tackles concept drifts effectively than the ARF + HAT.
Even though the model performance of HAT(10) was similar to the combined ARF + HAT, the results demonstrated that performance of the model alone does not necessarily indicate how an algorithm responds to concept drift.

Based on the experimental findings, the following conclusions can be drawn:
\begin{itemize}
\item  The results showed that nearly all ensemble techniques, both homogeneous and heterogeneous, significantly outperform single classification models in terms of Acc, R, P, and F1 performance measures,  although at the expense of increasing run-time.
\item Thus, based on our experiments, depending on specific application requirements and resources constraints, one can note by comparing models performance of the seven ensembles, that the best model can be either the heterogeneous ensemble (ARF + HAT) or the homogeneous ensemble HAT(10) , since both have delivered the highest predictive performance and overcome the misclassification of attacks in WSNs.
\item The online ensemble algorithm demonstrates great abilities to continuously process traffic data on a large scale because while the classifier learns, he attempts to train the model better, and the detection rate of attacks continues to improve over time/iterations as made evident by the improvement of recall, it is also applicable to any application, making it advantageous compared to existing models which are specific to their applications.
\end{itemize}

\section{Conclusion}
 Wireless sensor networks are often referred to as an emerging technology that will impact our daily lives. These electromechanical components of a very small size and which communicate via an ubiquitous wireless network, widely open the horizons of applications built up to now. Being exposed to numerous risks, the main challenge of the evolution of intrusion detection system in WSNs is to identify the attacks with great accuracy, and to respond to constraints and challenges required to extend the life of the entire network. Given that much more attention is paid to the detection techniques used, this goal could be achieved in a variety of ways.

 Our paper presents a novel perspective upon the malicious security attacks in WSNs by involving a fast intrusion detection scheme based on ensemble learning that satisfy the dynamic and continuous streaming of data. Our experiments with the WSN-DS attack database show that the ensemble approach performed better than any online classifier as an individual learner, despite having a generally longer run-time in distinguishing attacks from benign samples; thus, we propose heterogeneous ensemble (ARF + HAT) and homogeneous ensemble HAT(10), as both achieve higher detection rates in the aim of distinguishing malicious attacks while taking into account the resource constraints of WSNs when compared to other intrusion detection methods, and its prediction performance improves over time as new data points are integrated. In general, our proposed model is effective for real-time WSN intrusion detection.

In future work, we will explore different methods such as preprocessing with data reduction and parameter tuning, that can improve the efficiency of the classifier. The performance can be further improved using deep learning techniques to further enhance WSNs attack detection performance.

\bibliographystyle{unsrt}  
\bibliography{references}

\end{document}